\documentstyle[astrobib,psfig]{mn-ab}
\newcommand{\hmpc}{\mbox{$h^{-1}$ Mpc}}

\newcommand{\mpc}{\mbox{Mpc}}

\newcommand{\msun}{\mbox{$M_{\odot}$}}

\newcommand{\kmsmpc}{\mbox{km s$^{-1}$ Mpc$^{-1}$}}

\def\la{\mathrel{\hbox{\rlap{\hbox{\lower4pt\hbox{$\sim$}}}\hbox{$<$}}}}
\def\ga{\mathrel{\hbox{\rlap{\hbox{\lower4pt\hbox{$\sim$}}}\hbox{$>$}}}}

\title[Evaluating Semi-Analytic Halo Merging Histories]
	{Evaluating Semi-Analytic Halo Merging Histories}
\author[R.S. Somerville, G. Lemson, T.S. Kolatt \& A. Dekel]
       {Rachel S. Somerville$^{1,2}$, Gerard Lemson$^{1,3}$, 
	Tsafrir S. Kolatt$^{2}$ and Avishai Dekel$^1$\\
        $^1$Racah Institute of Physics, The Hebrew University, Jerusalem\\
        $^2$Physics Department, University of California, Santa Cruz \\
	$^3$Max-Planck Institut f\"{u}r Astrophysik, Garching}

\begin{document}

\maketitle

\begin{abstract}
We evaluate the accuracy of semi-analytic merger-trees by comparing them with
the merging histories of dark-matter halos in $N$-body simulations, focusing on
the {\it joint distribution} of the number of progenitors and their masses. We
first confirm that the halo mass function as predicted directly by the
Press-Schechter (PS) model deviates from the simulations by up to 50\% depending
on the mass scale and redshift, while the {\it means} of the projected
distributions of progenitor number and mass for a halo of a given mass are more
accurately predicted by the Extended PS model. We then use the full merger trees
to study the joint distribution as a function of redshift and parent-halo mass.
We find that while the deviation of the mean quantities due to the inaccuracy of
the Extended PS model partly propagates into the higher moments of the
distribution, the merger-tree procedure does not introduce a significant
additional source of error. In particular, certain properties of the merging
history such as the mass ratio of the progenitors and the total accretion rate
are reproduced quite accurately for galaxy sized halos ($\sim 10^{12}\msun$),
and less so for larger masses. We conclude that although there could be $\sim
50\%$ deviations in the absolute numbers and masses of progenitors and in the
higher order moment of these distributions, the {\it relative} properties of
progenitors for a given halo are reproduced fairly well by the merger
trees. They can thus provide a useful framework for modelling galaxy formation
once the above-mentioned limitations are taken into account.
\end{abstract}


\section{Introduction}
\label{sec:intro}
In the standard picture of hierarchical structure formation, small objects form
first and then merge together to form larger objects as time
progresses. Hierarchical structure formation is a generic prediction of the Cold
Dark Matter (CDM) family of models, being a natural consequence of the general
shape of the power spectrum. The process of structure formation can be
represented by the ``merging history'' of virialized dark matter halos, i.e. the
masses of the halos, identified at an earlier time when they were still distinct
entities, that will later merge together to form a larger halo. This merging
history is often referred to as a ``merger tree''. This whole conceptual
framework is somewhat artificial, relying as it does on a definition of
``halo'', which involves assumptions of sphericity, virialization, etc. These
properties almost certainly do not apply to all of the objects in the real
universe that we might like to study. However, one reason for its popularity is
that it enables one to develop analytic and semi-analytic predictions for many
quantities of interest.
 
For example, the model developed by \citeN{ps:74} provides a simple but
relatively effective framework for the description of the mass history of
particles in a hierarchical universe with Gaussian initial perturbations. The
original Press-Schechter (PS) model predicted the mass function of halos as a
function of redshift, i.e. the number density of halos of a given mass at a
redshift $z$. This prediction has been tested against the results of $N$-body
simulations in several previous papers
\cite{efstathiou:88,gelb:94,lc:94,ma:96,tozzi:97,gsphk:98,tormen:98}. The
Press-Schechter theory was extended to give the \emph{conditional} probability
that a particle in a halo of mass $M_0$ at $z_0$ was in a halo of mass $M_1$ at
an earlier redshift $z_1$, leading to an expression for the conditional halo
mass function \cite{bcek,bower:91}. This extended Press-Schechter (EPS)
formalism can also be manipulated to obtain expressions for halo survival
times, formation times, and merger rates \cite[hereafter LC93]{lc:93}. The
extended Press-Schechter model and its implications have not been quite as well
studied as the standard PS model, but \citeN{lc:94} and recently
\citeN{tormen:98} have presented comparisons of the EPS model with $N$-body
simulations.

However, the EPS formalism only provides us with \emph{mean} quantities. In
order to construct realizations of individual dark matter halo merging
histories, one must include some additional assumptions. Although several
methods for constructing merger trees have been proposed \cite{block1,kw,sk},
all of them involve some ad hoc ingredients. Such a method should be designed
to reproduce both the conditional mass function predicted by the EPS
model and to enforce conservation of mass at every stage. However, these
constraints are not necessarily consistent with one another. This leads to
certain ambiguities and subtleties in the process of constructing the merging
trees, which we will discuss later in this paper. For a fuller discussion of
these issues, however, see \citeN[hereafter SK]{sk}.

These merger trees form the framework for semi-analytic models of galaxy
formation such as those described by \citeN{kwg}, \citeN{cafnz}, and
\citeN{sp:98}. The properties of the galaxies created in these models
presumably depend to some degree on the merging history of the dark matter
halos. Most of the previous work using the semi-analytic models has focussed on
reproducing or predicting mean quantities and qualitative trends. However, it
would be useful to investigate whether the broader properties of the predicted
\emph{distribution} of model galaxies are consistent with observations. For
example, we might wish to investigate the scatter in well known observational
relations like the luminosity-linewidth relation, the color-magnitude relation,
or fundamental plane relations. But before we can trust the models for
evaluating these kinds of questions, we must make sure that the merger trees
not only satisfy the mean properties readily predicted by the EPS model, but
ideally also the full distribution function. This has not been thoroughly
investigated in previous work on this subject, and is the primary goal of this
paper. The PS and EPS formalism do not provide any information about this
distribution. Therefore we are forced to appeal to $N$-body simulations.

In this paper, we present a comparison between $N$-body simulations and
semi-analytic merger trees (SAM-trees), as well as the analytic predictions of
the EPS theory when they apply. We address several issues. First, we re-examine
the agreement of the direct predictions of the standard and extended
Press-Schechter model with the simulations. Although much of the previous work
on this topic has stressed the unexpectedly good agreement, we focus on the
discrepancies, their broader implications, possible causes and potential
resolutions. We then study the success of the merger-tree method in the
reconstruction of the distribution of progenitor number and mass compared to
the results of the simulations, and examine several statistics relevant to the
study of galaxy formation in a hierarchical context.

The outline of the paper is as follows. In \S\ref{sec:ps}, we give a brief
summary of the PS and EPS formalism. In \S\ref{sec:trees}, we summarize the
merger-tree method. In \S\ref{sec:sims}, we describe the $N$-body
simulations. In \S\ref{sec:comp}, we present the comparison between the
simulations and the SAM-trees. We discuss our results and conclude in
\S\ref{sec:conclusions}.

\section{The Press-Schechter Formalism}
\label{sec:ps}
Suppose that we have smoothed the initial density distribution on a scale $R$
using some spherically symmetric window function $W_{M}(r)$, where $M(R)$ is
the average mass contained within the window function. There are various
possible choices for the form of the window function (c.f. \citeNP{lc:93}). We
use a real-space top-hat window function, $W_M(r) = \Theta(R-r)(4\pi
R^3/3)^{-1}$, where $\Theta$ is the Heaviside step function. In this case $M =
4\pi \rho_0 R^3/3$, where $\rho_0$ is the mean mass density of the
universe. The mass variance $S(M) \equiv \sigma^2(M)$ may be calculated from
\begin{equation}
\label{eqn:massvariance}
\sigma^2(M) = \frac{1}{2\pi^2} \int P(k) W^2(kR) k^2 {\rm d}k \, ,
\end{equation}
where $P(k)$ is the power spectrum of the matter density fluctuation, and
$W(kR)$ is the Fourier transform of the real space top-hat.

The ``excursion set'' derivation due to \citeN{bcek} leads naturally to the
extended Press-Schechter formalism. The smoothed field $\delta(M)$ is a
Gaussian random variable with zero mean and variance $S$. The value of $\delta$
executes a random walk as the smoothing scale is changed. Adopting an ansatz
similar to that of the original Press-Schechter model, we associate the
fraction of matter in collapsed objects in the mass interval $M, M+{\rm d}M$ at
time $t$ with the fraction of trajectories that make their \emph{first
upcrossing} through the threshold $\omega \equiv \delta_c(t)$ in the interval
$S, S+{\rm d}S$. This may be translated to a mass interval through equation
(\ref{eqn:massvariance}). The threshold $\delta_c(t)$ corresponds to the
critical density at which a pertubation will separate from the background
expansion, turn around, and collapse. It is extrapolated using linear theory,
$\delta_c(t)=\delta_{c,0}/D_{\rm lin}(z)$, where $D_{\rm lin}(z)$ is the linear
growth function \cite{peebles:80}.  The halo mass function (here in the
notation of LC93) is then:
\begin{equation}
\label{eqn:ps}
f(S, \omega) {\rm d}S = \frac{1}{\sqrt{2\pi}} \frac{\omega}{S^{3/2}} 
\exp{\left[-\frac{\omega^2}{2S}\right]} {\rm d}S \, .
\end{equation}

The \emph{conditional} mass function, the fraction of the trajectories in halos
with mass $M_1$ at $z_1$ that are in halos with mass $M_0$ at $z_0$ ($M_1 <
M_0$, $z_0 < z_1$) is 
\begin{eqnarray}
\label{eqn:flc}
\lefteqn{f(S_1, \omega_1 \mid S_0, \omega_0) {\rm d}S_1 =} 
\nonumber \hspace{1truecm}\\
 & &  \frac{1}{\sqrt{2\pi}} \frac{(\omega_1-\omega_0)}{(S_1-S_0)^{3/2}}
\exp{\left[-\frac{(\omega_1-\omega_0)^2}{2(S_1-S_0)}\right]} {\rm d}S_1
\,.
\end{eqnarray}
The probability that a halo of mass $M_0$ at redshift $z_0$ had a
progenitor in the mass range $(M_1, M_1+{\rm d}M_1)$ is given by (LC93):
\begin{eqnarray}
\label{eqn:Nlc}
\lefteqn{\frac{{\rm d}P}{{\rm d}M_1}(M_1, z_1 \mid M_0, z_0) {\rm d}M_1 =}
\nonumber \hspace{2.5truecm}\\
& & \frac{M_0}{M_1} f(S_1, \omega_1 \mid S_0, \omega_0) 
\left| \frac{{\rm d}S}{{\rm d}M}\right| {\rm d}M_1\, ,
\end{eqnarray}
where the factor $M_0/M_1$ converts the counting from mass weighting to number
weighting.

We can now derive two more quantities that will be useful later. Given the mass
of a parent halo $M_0$ and the redshift step $z_0 \rightarrow z_1$, the
\emph{average} number of progenitors with mass larger than $M_l$ is:
\begin{equation}
\label{eqn:avgnumprog}
\bar{N} \equiv
\langle N_p (M \,\vert M_0) \rangle = \int_{M_l}^{M_0} \, {\rm d}M\,
\frac{M_0}{M}\, \frac{{\rm d}P}{{\rm d}M}(M,z_1 \vert M_0, z_0) 
\,.
\end{equation}
We can also readily calculate the average fraction of $M_0$ that dwelt
in the form of progenitor halos of mass $M>M_l$:
\begin{equation}
\label{eqn:faccbar}
\bar{f}_p =\int_{M_l}^\infty{\rm d}M\, \frac{{\rm d}P}{{\rm d}M}
(M,z_1 \vert M_0,z_0)\,,
\end{equation}
and define the complimentary quantity for the average fraction of $M_0$ that
came from ``accreted'' mass, $\bar{f}_{\rm acc} =1-\bar{f}_p$.

Throughout this paper we use the standard value for the collapse threshold,
$\delta_{c,0}=1.69$, corresponding to the value predicted by the top-hat
spherical collapse model.

\section{The Trees}
\label{sec:trees}
We first should clarify some of our terminology. Suppose we have identified a
halo of a given mass at some redshift, whose merging history we wish to
know. This halo is referred to as the ``parent''. We then go backwards in time
and trace all the smaller halos that merged together in order to form this
halo. These are referred to as ``progenitors''. Although it violates the usual
temporal relation of kinship, because we construct the tree working backwards
in time, this has become the usual terminology.

Let us suppose that we set out to construct a merger tree, taking the EPS
formalism as an ansatz. Such a merger tree should satisfy at least two basic
requirements: each individual halo history should conserve mass at every stage,
and the EPS and PS mass functions (for every redshift) should be
reconstructed in the mean for a large ensemble of realizations. Unfortunately,
satisfying both requirements exactly and simultaneously is not straightforward,
and is sometimes impossible for certain choices of power spectrum and redshift
interval. There are also additional problems -- the EPS formalism gives us the
\emph{single} halo probability that a halo with mass $M_0$ at $z_0$ has a
progenitor of mass $M_1$ at $z_1$ --- it does not provide us with any
information about the rest of the progenitors that make up the mass of $M_0$
(i.e. the \emph{joint} progenitor probability function). There are certain
obvious constraints which the EPS formalism knows nothing about, for example,
the sum of the progenitor masses clearly cannot exceed $M_0$. In addition,
since the number of halos diverges at very small mass, one must impose some
minimum mass $M_l$, below which one no longer follows the history of these
halos. However, one must account for the mass contributed by these small
halos. We will refer to halos with masses larger than $M_l$ as ``progenitors'',
and the mass contributed by halos smaller than this limit as ``accreted mass''.

One must therefore form a compromise between mass conservation and
reconstructing the mean as predicted by the EPS model, and one must make some
ad-hoc assumptions in order to fill in the gaps mentioned above. In the method
of \citeN{kw}, the mean progenitor mass distribution is reproduced exactly, and
mass conservation is enforced only approximately. In the scheme of \citeN{sk},
mass conservation is enforced exactly and the mean progenitor distribution is
reproduced approximately.

We now sketch the method of SK. Given a parent halo with mass $M_0$, the
histories of individual particles, or trajectories, in the language of the
excursion set formalism, are followed back in time using very small
timesteps. For each particle that finds itself in $M_0$ at $z_0$, we choose the
mass of the progenitor halo it was located in at an earlier redshift $z_1$ from
Eqn.~\ref{eqn:flc}. A ``branch'' occurs when at least two trajectories are
found in halos with masses greater than the minimum halo mass $M_l$. The
timestep has been chosen such that ``branching'' events are improbable; i.e. in
most cases one trajectory will be in a halo with mass close to $M_0$ and others
will be in halos with masses less than $M_l$. This ensures that most of the
branching events involve small numbers of branches, but no limit is imposed on
the number of branches allowed. In order to avoid choosing a progenitor with a
mass larger than the remaining unallocated mass, the probability function is
truncated for masses larger than this value. The contribution from ``accreted
mass'' is included explicitly from trajectories that find themselves in halos
smaller than $M_l$. It is possible to include these without introducing a large
computational overhead because the \emph{mass weighted} probability, used for
the trajectories, does not diverge at small mass. One continues to follow
trajectories until all of the mass of the parent has been accounted
for. Therefore mass is conserved exactly. However, the shape of the conditional
mass function that is obtained, although very similar to the EPS prediction we
wish to reconstruct, is not exact (see SK).

\section{The Simulations}
\label{sec:sims}
In order to study the formation histories of halos in numerical simulations, we
need:
\begin{enumerate}
\item High mass and force resolution

\item A large enough volume to contain many halos in the mass range of interest

\item Stored data for several time steps, at appropriate redshifts.
\end{enumerate}
We use the adaptive particle-particle/particle-mesh simulations of the VIRGO
consortium \cite{virgo}. The resolution and box size of the
smaller volume simulations in the suite are appropriate for our purposes, since
we wish to focus on galaxy or group sized halos. 
 
We present our results for the $\tau$CDM model, a critical ($\Omega=1$) model
with shape parameter $\Gamma = 0.21$. This value of the shape parameter is
motivated by observations of galaxy clustering
\cite{maddox:90,maddox:96,vogeley:92}. The parameters of the simulation are
given in Table~\ref{tab:models}. We have performed the same analysis for a
simulation with $\Omega_0=0.3$ and an open geometry, and we find qualitatively
similar results to those we will describe for the $\tau$CDM model, leading to
similar conclusions. We therefore do not present the results from other
cosmological models.

\begin{table*}
\caption{Simulation Parameters. From left to right, the quantities given are
the mass density in units of the critical density, the Hubble parameter
$h\equiv H_0/(100\, \kmsmpc)$, the linear rms mass variance on a scale of $8
\hmpc$, the number of particles, the number of grid cells, the length of one
side of the box, and the mass per particle. }
\begin{center}
\begin{tabular}{lccccccc}
\hline 
Model  & $\Omega_0$ & $h$ & $\sigma_8$ & $N_p$ & $N_g$ & $L_{\rm box}$
(\mpc) & $m_{p}$ (\msun)\\
\hline
$\tau$CDM & 1.0 & 0.5 & 0.6 & $256^3$ & 512 & 170 & $2.0\times10^{10}$\\
%OCDM & 0.3 & 0.7 &0.85 & $256^3$ & 512 & 141 & $1.4\times10^{10}$  \\
\hline
\end{tabular}
\label{tab:models}
\end{center}
\end{table*}

The halos are identified using a standard ``friends-of-friends'' algorithm with
a linking length $b=0.2\bar{d}$, where $\bar{d}$ is the mean interparticle
spacing $\bar{d} = L_{\rm BOX}/256$. The friends-of-friends algorithm with this
value for the linking length has been shown to give roughly equivalent results
to spherical over-density based halo finders \cite{lc:94}.

The procedure to identify progenitors and assign their masses is as
follows. Consider a parent halo with mass $M_0$ at $z=0$. We track each
particle in this halo to an earlier redshift, $z_1$, and find the mass of the
halo in which it is located at that time. We then identify all the other
particles in the progenitor halo that are also present in the parent halo $M_0$
at $z=0$. At this point, we may be faced with several problems which reflect
the inconsistency of the simple merging hierarchy picture with the numerical
properties of $N$-body simulations. We may encounter the unphysical situation
where a progenitor halo identified in this way has a mass larger than the
parent. This is simply because individual particles are often stripped or
ejected from halos in the simulations. If this occurs, we exclude the parent
halo from our sample. In addition, not all the particles in the progenitor end
up in the same parent halo. We then have two possible choices of ways to assign
mass to the progenitor. We can either take only the mass of all the particles
in the progenitor that also end up in the parent halo, or we can take the total
mass of the progenitor halo. In the first case, the mass does not reflect the
dynamical mass of the progenitor, which is the quantity that is treated by the
PS formalism. In the second case, conservation of mass is violated. Since it is
not clear which is the most consistent prescription, and since these two
choices in some sense represent opposite extremes, we have performed our
analysis using both schemes. We shall refer to the masses identified in these
two ways as the ``inclusive'' (mass of particles that end up in parent) and
``total'' (total virial mass of progenitor) progenitor masses.

Recall that in the SAM-tree procedure, we defined a minimum progenitor mass
$M_l$. In the case of the trees, this mass scale is arbitrary, but in the
simulations, we should choose it to correspond to the smallest halo that we
believe to be numerically stable. We have tried using a minimal halo mass
corresponding to both 10 and 25 particles. We show the results for the choice
of 25 particles because the agreement between Press-Schechter and the
simulations is better, and because the results for the two different
definitions of progenitor mass discussed above are more consistent with one
another. This leads us to believe that the results are less affected by the
numerical instability of the halos if we only consider halos of more than 25
particles. As in the trees, we now term the total amount of mass contributed by
halos with masses smaller than $M_l$ as the ``accreted mass''. The estimate of
the accreted mass also depends on the method used to assign progenitor
masses. For the ``inclusive'' scheme, the mass of the halo is more likely to
fall below $M_l$ and hence the estimate of accreted mass tends to be larger.

\section{The Comparison}
\label{sec:comp}
\subsection{Mean Quantities}
\label{sec:comp:means}
In this section, we compare the results of the simulations described in the last
section with the predictions of the PS and EPS models, and with the results of
the SAM-trees described in \S\ref{sec:trees}.  In the remainder of the paper, we
will express all masses in terms of the minimum progenitor mass, $M_l$, which
here corresponds to a physical mass of $25 \, m_p = 5.0 \times 10^{11} \msun$.
In order to have sufficient statistics, we group the parent halos in the
simulations into ten logarithmically spaced mass bins, spanning the mass range
from $5 \, M_l$ to $500 \, M_l$.  We generate the SAM-trees for a grid of parent
masses covering the range of $M_0$ in each mass bin used for the simulations,
weighting the contribution according to the Press-Schechter number density at
$z=0$. This is an attempt to mimic the range of parent masses represented in a
given bin in the simulations, and accordingly any scatter in the distributions
that may result from this effect.

\begin{figure}
\centerline{\psfig{file=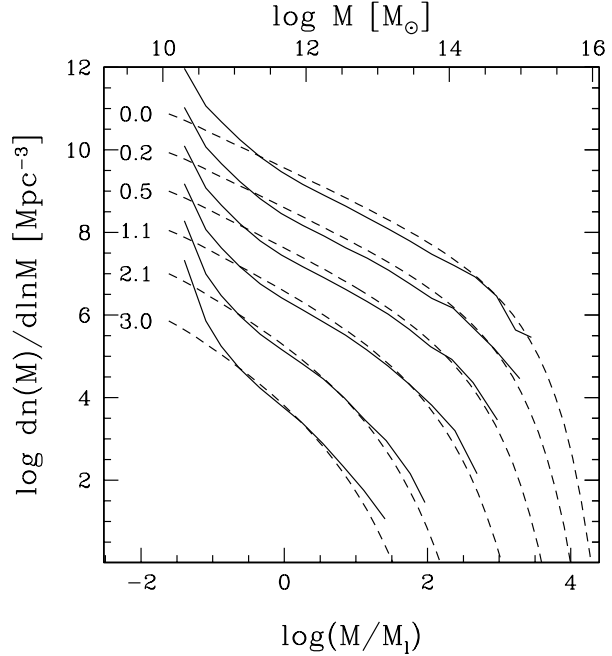,height=9truecm,width=9truecm}}
\caption{The mass function of halos predicted by the standard Press-Schechter
model (dashed lines), and found in the simulations (solid lines) at redshifts
of (from top to bottom) $z=0$, 0.2, 0.5, 1.1, 2.1, and 3.0.}
\label{fig:mf}
\end{figure}
Figure~\ref{fig:mf} shows the overall halo mass function for the simulations,
compared with the standard Press-Schechter model. At $z=0$, the PS model agrees
well with the simulations at high masses ($\ga 10^{14} \msun$), and overpredicts
the number of halos by about a factor of two between masses of $10^{14}$ and
$\sim 5 \times 10^{11} \msun$. At very small masses ($\la 2\times10^{11}\msun$),
the simulations have many more halos than the PS model prediction, which is
almost certainly due to numerical effects. 

The scatter in the simulation mass function due to shot noise is insignificant
on the mass scales where the discrepancy with the Press-Schechter model has been
noted (at $z=0$, less than 5 percent for $M \la 10^{13} \msun$, less than 10
percent for $M \simeq 10^{14}
\msun$). If plotted as error bars on Fig.~\ref{fig:mf}, the width of the error
bars on these scales would be about the same as the width of the line used to
plot the mass function. The shot noise becomes significant (greater than 20
percent) only for the two largest mass bins ($M > 5 \times 10^{14}$), which
contain only a few halos because of the limited volume of our simulation box.

The mass scale at which the PS model starts to underestimate the number of halos
moves to smaller values with increasing redshift. This means that by a redshift
of $z\sim 3$, the simulations have considerably more halos than the PS
prediction at masses $\ga 10^{12} \msun$; i.e. the evolution of the mass
function with redshift is exaggerated in the PS model. However, the mass scale
of the steepening on the small mass end remains constant with redshift, as
expected if it is indeed due to numerical effects.

It should be noted that \citeN{gsphk:98} and \citeN{mike-thesis} find the
similar results (both the nature and magnitude of the discrepancy and its
dependence on redshift) using a different kind of code (Particle-Mesh) and a
different method for identifying halos (spherical and elliptical overdensity
methods). The same results are obtained in all five of the CDM-variant
cosmologies simulated by
\citeN{gsphk:98}. Similar results have also been obtained by \citeN{tormen:98}
and \citeN{tozzi:97}. It should also be noted that changing the value of the
parameter $\delta_{c,0}$ will only cause the PS mass function to cross the
simulation mass function at a different mass scale and cannot bring the mass
functions into good agreement simultaneously on all scales.

\begin{figure}
\centerline{\psfig{file=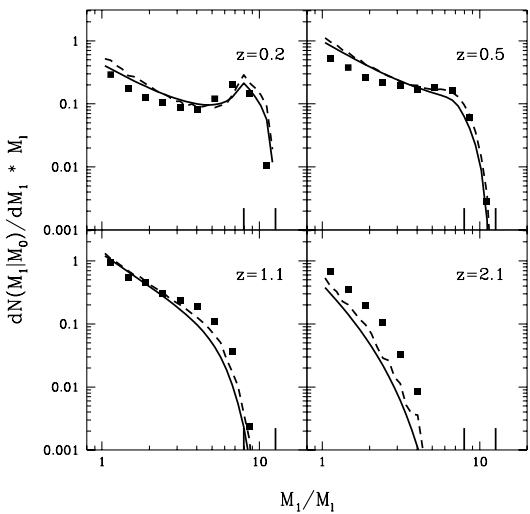,height=8truecm,width=8truecm}}
%\centerline{\psfig{file=conmult40.ps,width=\columnwidth}}
\caption{The conditional mass function of halos predicted by the Extended
Press-Schechter model (solid lines), found in the simulations (filled squares),
and in the SAM-trees (dashed lines). The Extended Press-Schechter prediction
and the SAM-trees have been averaged over the bin in parent mass used for the
simulations, indicated by the vertical lines on the bottom of the box. The
panels show redshifts 0.2, 0.5, 1.1, and 2.1 as indicated on the figure.
%The top panel shows a mean parent halo mass of $\bar{M}_0=10 M_l$ and 
%the bottom panel shows $\bar{M}_0=40 M_l$ 
The mean parent mass is $\bar{M}_0=10 M_l$.
}
\label{fig:cmf}
\end{figure}
Figure~\ref{fig:cmf} shows the conditional mass function for a bin in
parent halo mass $M_0=7.9-12.6 \,M_l$; i.e. the probability that the parent
halo had a progenitor of mass $M_1$ at some earlier redshift $z_1$.
% and $M_0=31.5-50.0\, M_l$. 
The geometric mean parent mass is approximately $10 \,M_l$, which 
%and $40\,M_l$. 
with our choice of $M_l=5.0\times10^{11}$, corresponds to the mass
of the halo surrounding an $L_{*}$ galaxy ($M=5.0\times10^{12}\msun$).
%and asmall group ($M = 2\times10^{13}$).
In addition to being a relevant mass scale if we are interested in studying
galaxy formation, there are many such halos in our simulation volume, so that
shot noise is insignificant for the remainder of the comparisons that we shall
discuss.  Figure~\ref{fig:cmf} shows the prediction of the EPS theory
(Eqn.~\ref{eqn:Nlc}), the SAM-trees, and the results of the simulations. The
EPS prediction and the SAM-trees have been averaged over the bin in parent halo
mass used for the simulations. We find that the simulation results are reasonably
similar for the two progenitor mass identification schemes described in
Section~\ref{sec:sims}, so we simply average the values obtained in the two
schemes for the remainder of the paper.

Note that the discrepancy between the EPS theory and the simulations is
qualitatively similar to the discrepancy that we saw before between the
standard PS and the total mass function. The simulations have fewer halos than
EPS in the intermediate mass range, and (though this is not shown in the
figure) steepen at very small masses ($\la M_l$). We also see the same reversal
in the trend as a function of redshift. The peak near $M_0$ is moved to
somewhat smaller masses. This means that the largest progenitor in the
simulations is somewhat smaller than in the the EPS-based trees, and the
complementary accreted mass is larger. Similar effects are seen for
cluster-sized halos in the simulations of \citeN{tormen:98}.

\subsection{Distributions}
\label{sec:comp:dist}
\begin{figure}
\centerline{\psfig{file=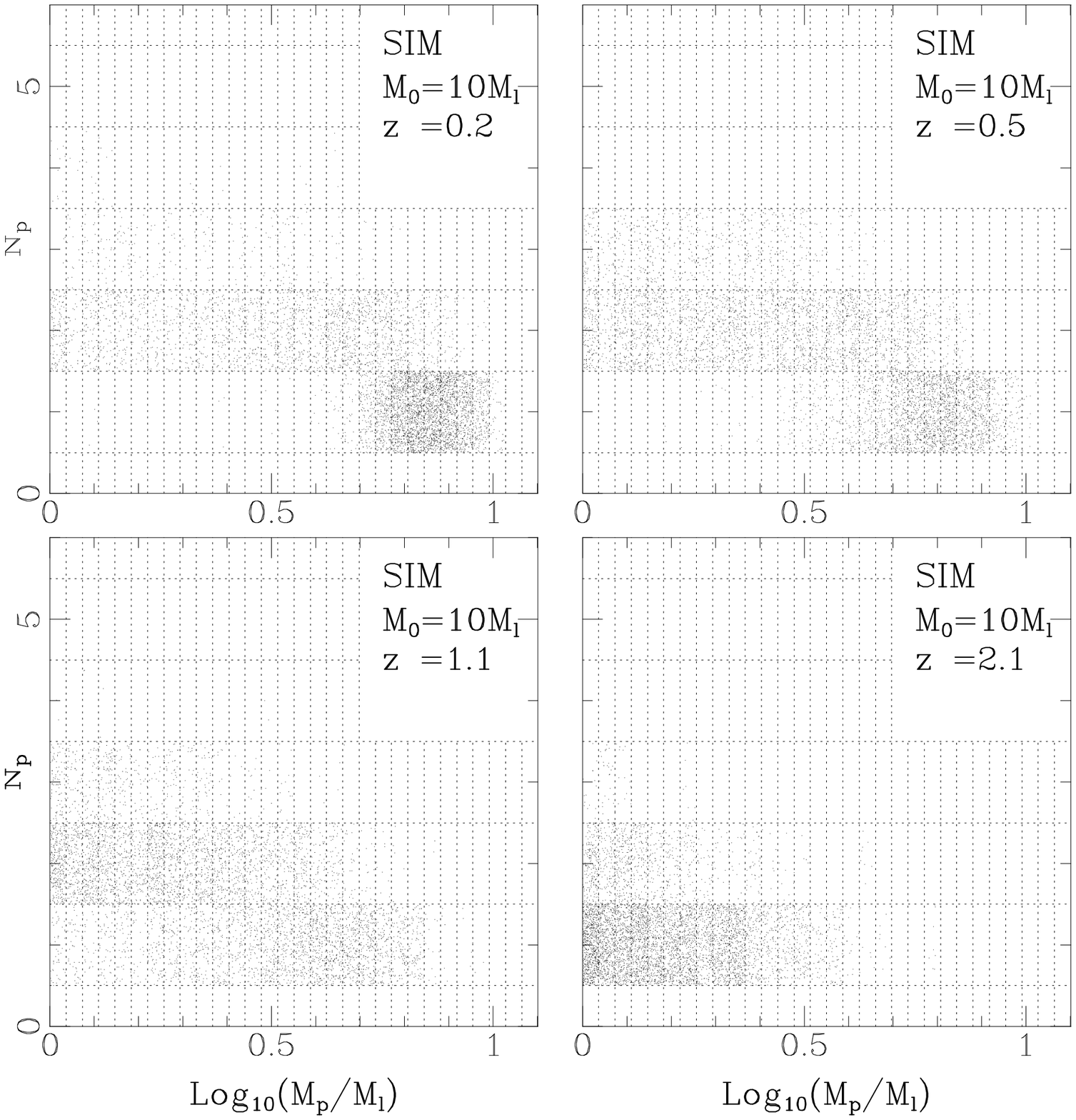,height=8truecm,width=8truecm}}
\centerline{\psfig{file=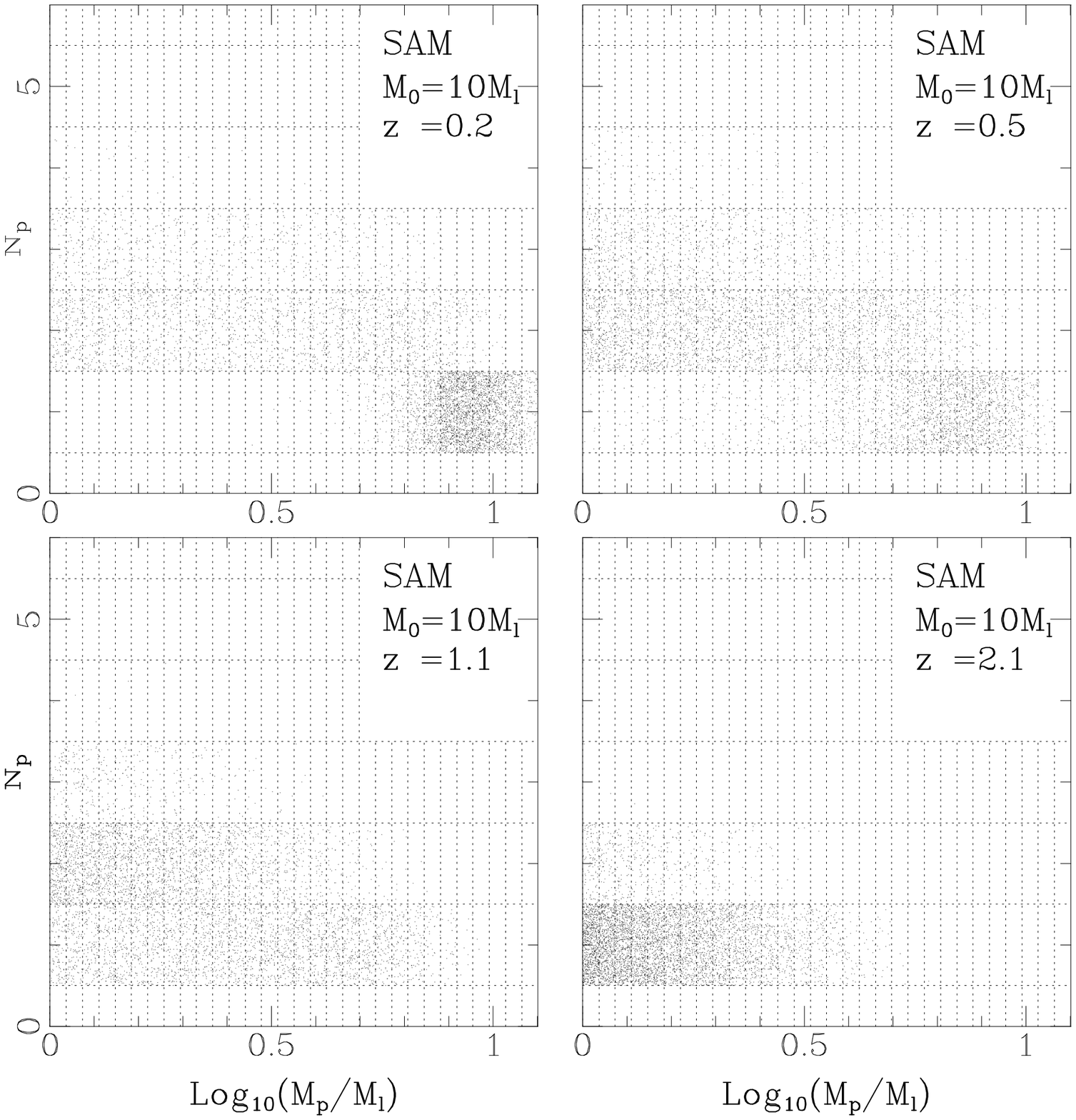,height=8truecm,width=8truecm}}
\caption{The two-dimensional distribution of the number of progenitors and
their masses at redshifts 0.2, 0.5, 1.1, and 2.1, for parent halos with average
mass $\bar{M}_0=10\,M_l$. The points have been plotted with random offsets
within the bins. The top panel shows the simulations, and the bottom panel show
the SAM-trees.}
\label{fig:progtrunk_10}
\end{figure}

\begin{figure}
\centerline{\psfig{file=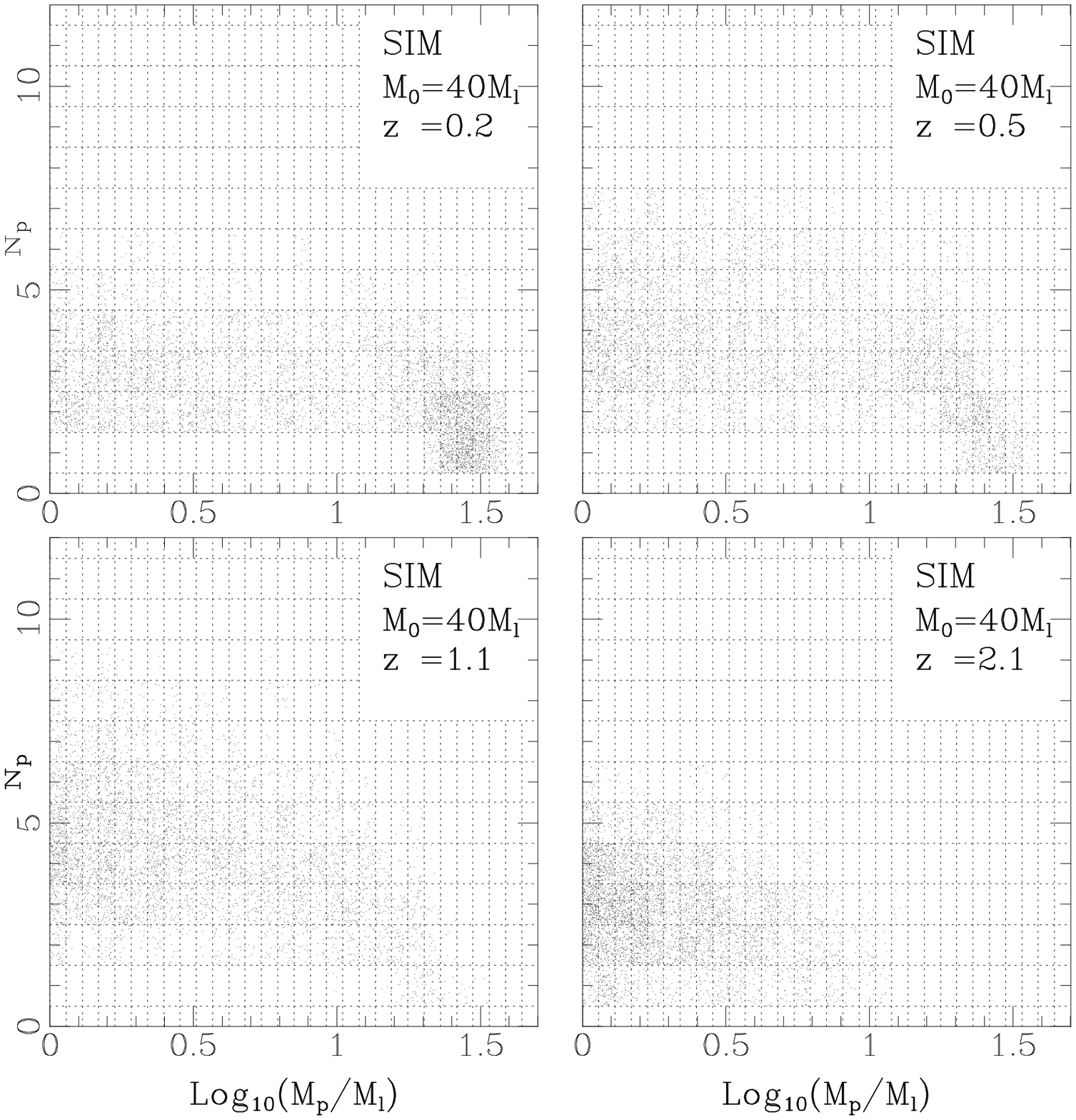,height=8truecm,width=8truecm}}
\centerline{\psfig{file=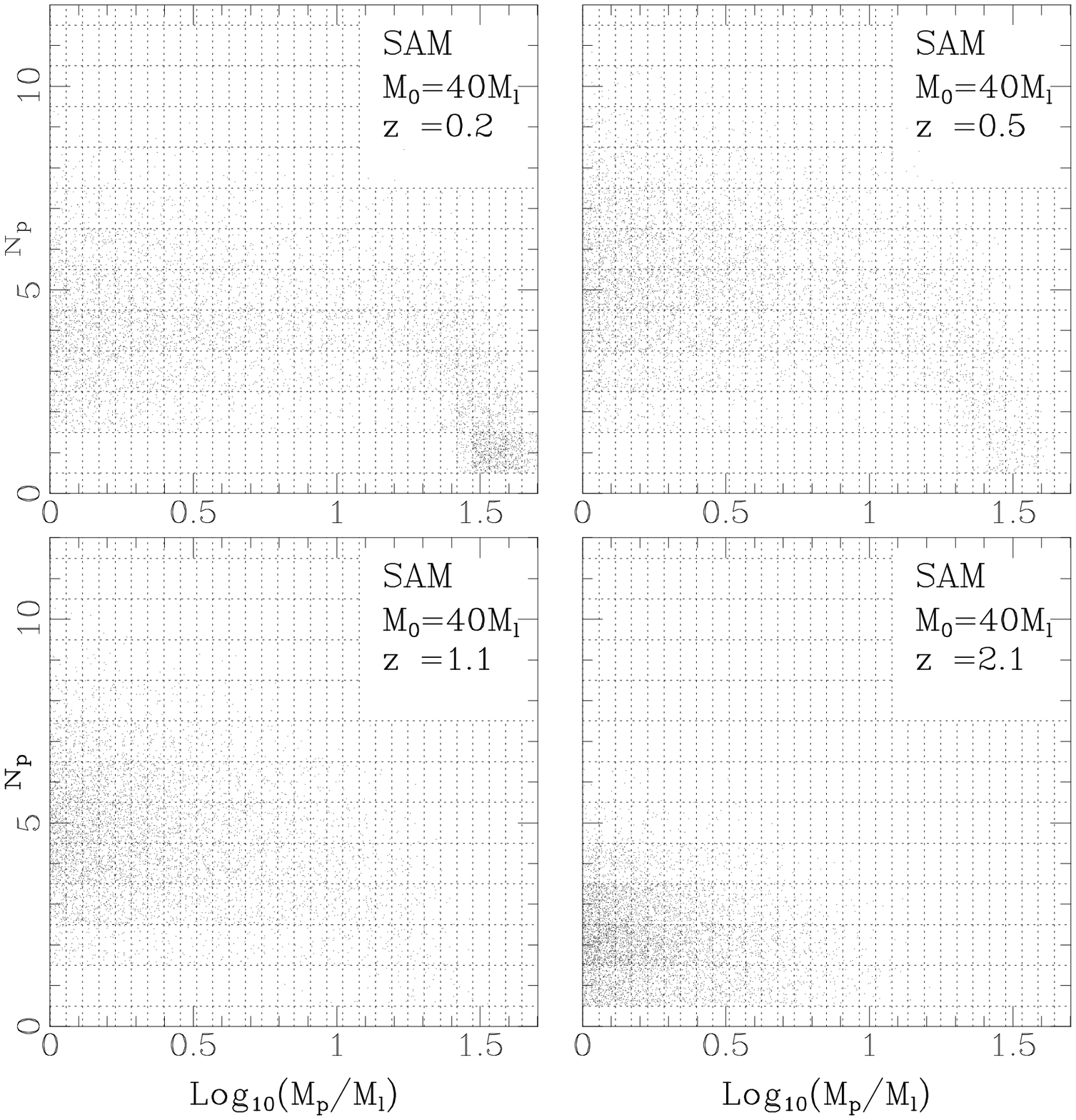,height=8truecm,width=8truecm}}
\caption{Same as figure \ref{fig:progtrunk_10} for $\bar{M}_0=40\,M_l$.}
\label{fig:progtrunk_40}
\end{figure}

We now address the distribution of progenitor mass and number, i.e. the
probability that a halo of mass $M_0$ at $z_0$ had a progenitor at $z_1$ with
mass $M_p$, and that this progenitor was a member of a set of $n$
progenitors. We denote this as $P(n, M_p)$. We show this
distribution in figures~\ref{fig:progtrunk_10}--\ref{fig:progtrunk_40}. The
number of points plotted in each bin is proportional to $P(n, M_p)$. We have
offset the points randomly within the bins for clarity, although $n$ can
actually take only integer values. We show the results for parent halos in mass
ranges centered on $\bar{M}_0=10\,M_l$ and $\bar{M}_0=40\, M_l$, for both the
simulations and the SAM-trees. Note that the smaller mass parent corresponds to
the mass of the dark matter halo surrounding an isolated $L_{*}$ galaxy, and
the larger mass parent to the dark matter halo surrounding a group.

These distributions show qualitatively the same behaviour for the SAM-trees and
the simulations. At $z=0.2$ most of the progenitors are single progenitors with
masses close to that of the parent, with the rest of the mass coming from
accretion (halos smaller than $M_l$). As we go back to progressively higher
redshifts, the mean progenitor mass shifts towards smaller masses, but the mean
number of progenitors stays roughly constant. With some imagination, one can see
the shape of an elephant head in these plots: the wide roundish part on the left
is the ``head'' and the curved plume extended out to the right is the ``trunk''
(especially noticable in the larger mass parent halo shown
(figure~\ref{fig:progtrunk_40})). At low redshift, there is a clump of
progenitors with mass close to $M_0$ and small numbers of progenitors, which
broadens as it moves towards larger numbers of smaller progenitors. As we move
to higher redshift, this clump at the end of the ``trunk'' drains further back
into the ``head'', and the trunk becomes less and less pronounced, until by
$z=2$ there is almost no trace of it. The ``trunkiness'' (i.e. the presence of
the thin extended plume) persists longer (until higher redshift) for larger mass
parents.  Elephants with trunks correspond to a situation where halos are formed
from several progenitors, and in this case we would expect to find several
galaxies within a single dark matter halo. Elephants without trunks (the more
rounded distributions) correspond to a situation where there is often a single
progenitor that has mostly grown by accreting mass. Thus in a very general way,
we can understand from these distributions how the merging history of an
isolated $L_{*}$ galaxy differs from one living within a group or cluster. This
characteristic shape is a direct result of the combined constraint of the
progenitor mass distribution (figure~\ref{fig:cmf}) and mass conservation. Thus
any merger tree method that satisfies these constraints should reproduce the
general properties of the mass-number distribution, regardless of the details of
the method.

When we try to compare the distributions more stringently, we immediately
notice several differences. In the first redshift interval, the SAM-trees have
progenitors filling in all the mass bins up until the largest bin. This
indicates that there are progenitors with masses very close to $M_0$, which
have accreted a very small amount of mass. In the simulations, the largest
progenitor mass bins are almost empty. This is a reflection of the same
discrepancy that we saw in the conditional mass function
(figure~\ref{fig:cmf}) --- the peak in progenitor mass near $M_0$ was shifted
towards smaller masses in the simulations. This means that the largest
progenitor is smaller and the average accreted mass is larger in the
simulations. Another, related difference in the $P(n, M_p)$ distributions is
that in the SAM-trees, the ``elephant head'' is more diffuse and is shifted
towards larger numbers of progenitors. The elephants also lose their trunks
faster as one goes to higher redshift in the SAM-trees (indicating that the
evolution is faster). This again mirrors what we saw in the conditional and
overall mass functions: the simulations have more high-mass halos at
high redshift than the EPS model predicts, indicating less evolution. These
discrepancies become more and more noticeable for larger parent masses. We
therefore see that the discrepancies in the mean quantities predicted directly
by the PS and EPS theory compared with the simulations lead to discrepancies in
the merging history of the halos obtained from the EPS-based SAM-trees.

\begin{figure}
\centerline{\psfig{file=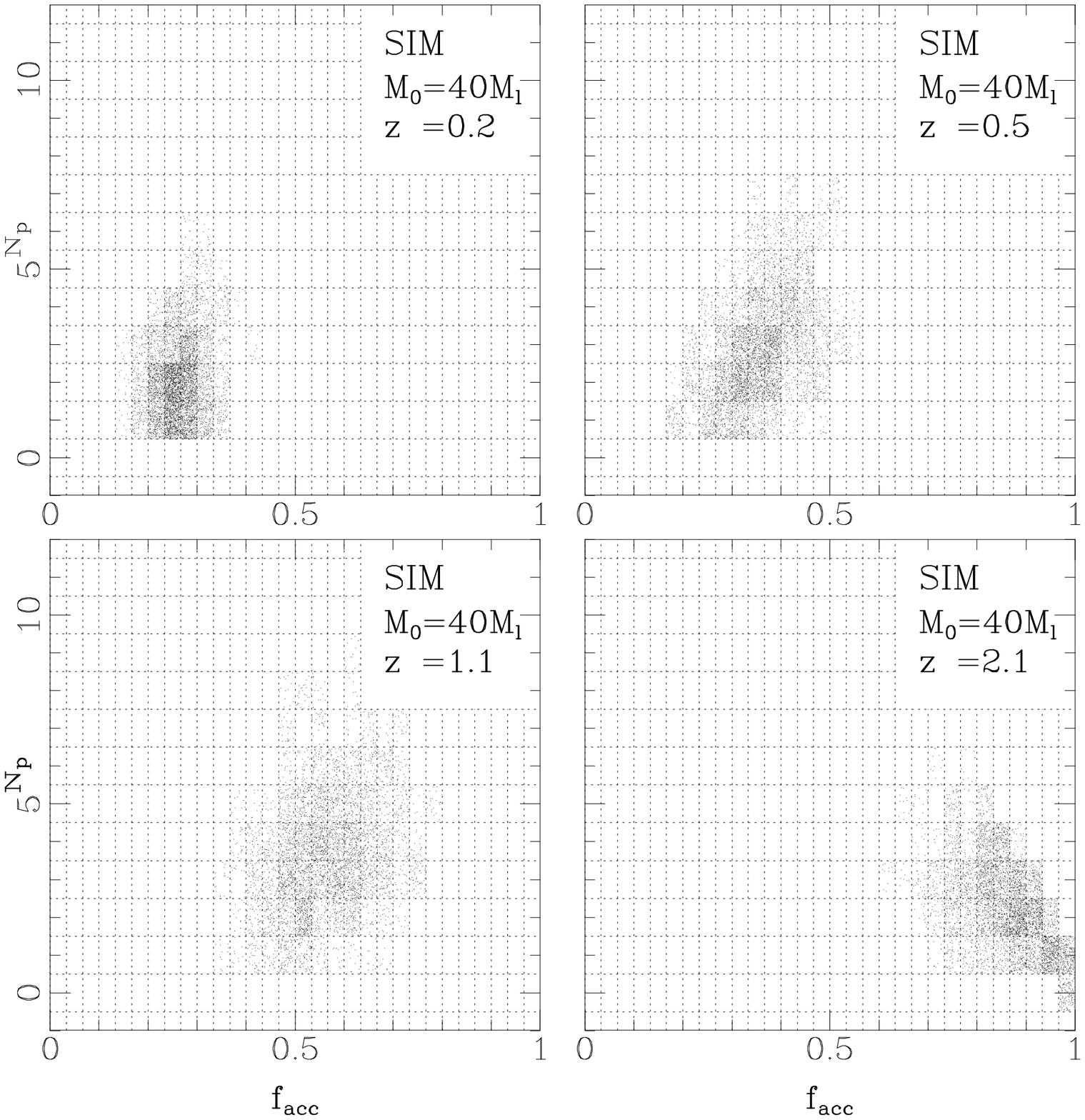,height=8truecm,width=8truecm}}
\centerline{\psfig{file=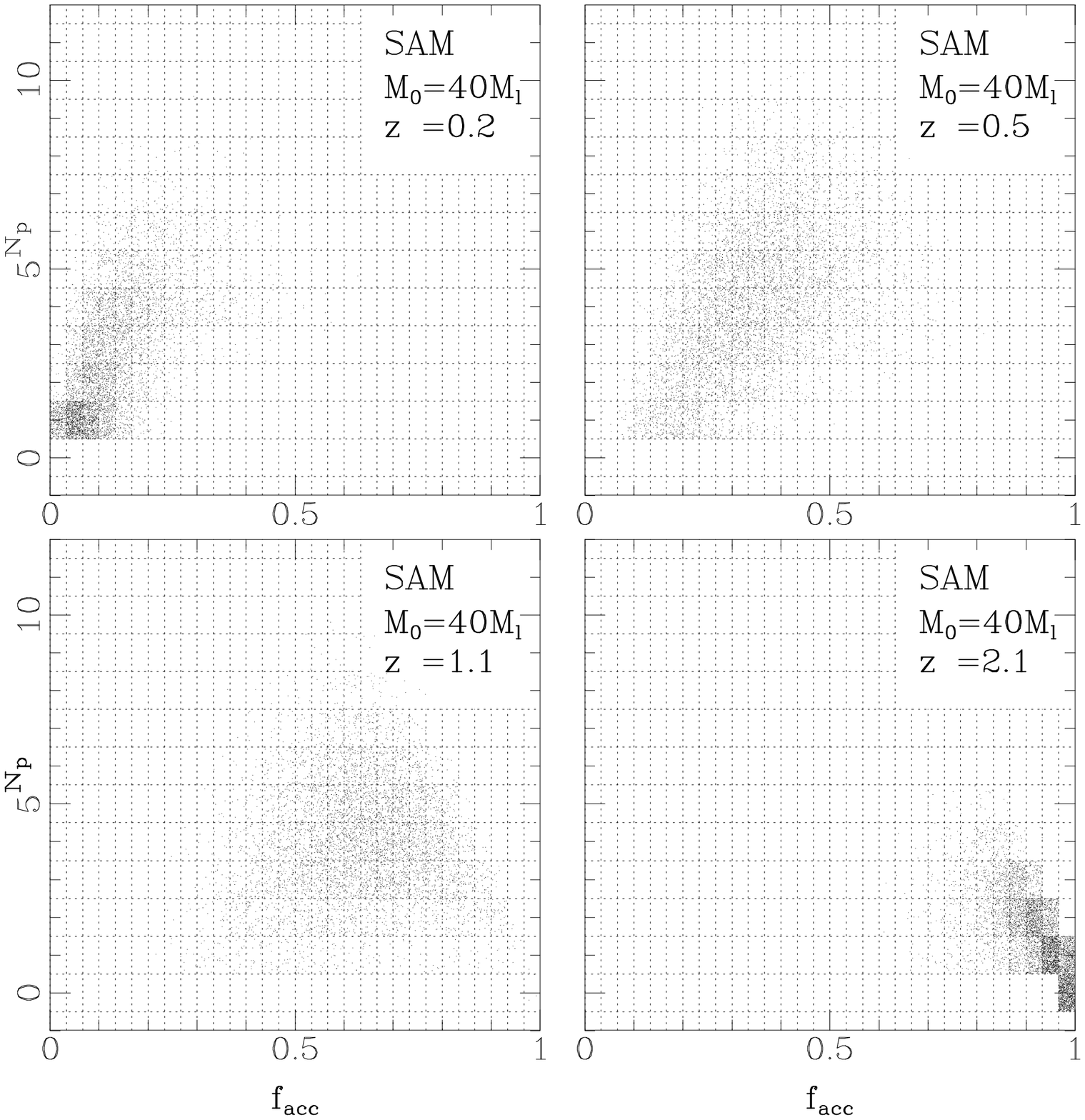,height=8truecm,width=8truecm}}
\caption{The two-dimensional distribution of the number of progenitors and the
fraction of the parent mass in the form of accreted mass, at redshifts 0.2,
0.5, 1.1, and 2.1, for parent halos with average mass $\bar{M}_0=40\,M_l$. The
points have been plotted with random offsets within the bins. The top panel
shows the simulations, and the bottom panel shows the SAM-trees.}
\label{fig:facctrunks_40}
\end{figure}
The complement of the quantity presented above is the distribution of
progenitor number and accreted mass, $P(n, f_{\rm acc})$. This is the
probability that a halo with mass $M_0$ had $n$ progenitors and $f_{\rm acc}
\equiv M_{\rm acc}/M_0$ in accreted mass. In figure~\ref{fig:facctrunks_40}, we
indeed see a behaviour that is complementary to that we have just discussed
with relation to the distribution $P(n, M_p)$. The fraction of the parent mass
in the form of accreted mass is larger in the simulations, and there is less
scatter in both $f_{\rm acc}$ and $N_p$ than in the SAM-trees.  The accreted
mass increases as we move to higher redshift, reflecting the fact that a larger
fraction of the mass in the universe is in the form of very small halos. The
number of progenitors increases until a redshift of about $z=1$, when the
number decreases as more of the progenitors fall below the mass resolution.

\subsection{Projections}
\label{sec:comp:proj}
\begin{figure}
\centerline{\psfig{file=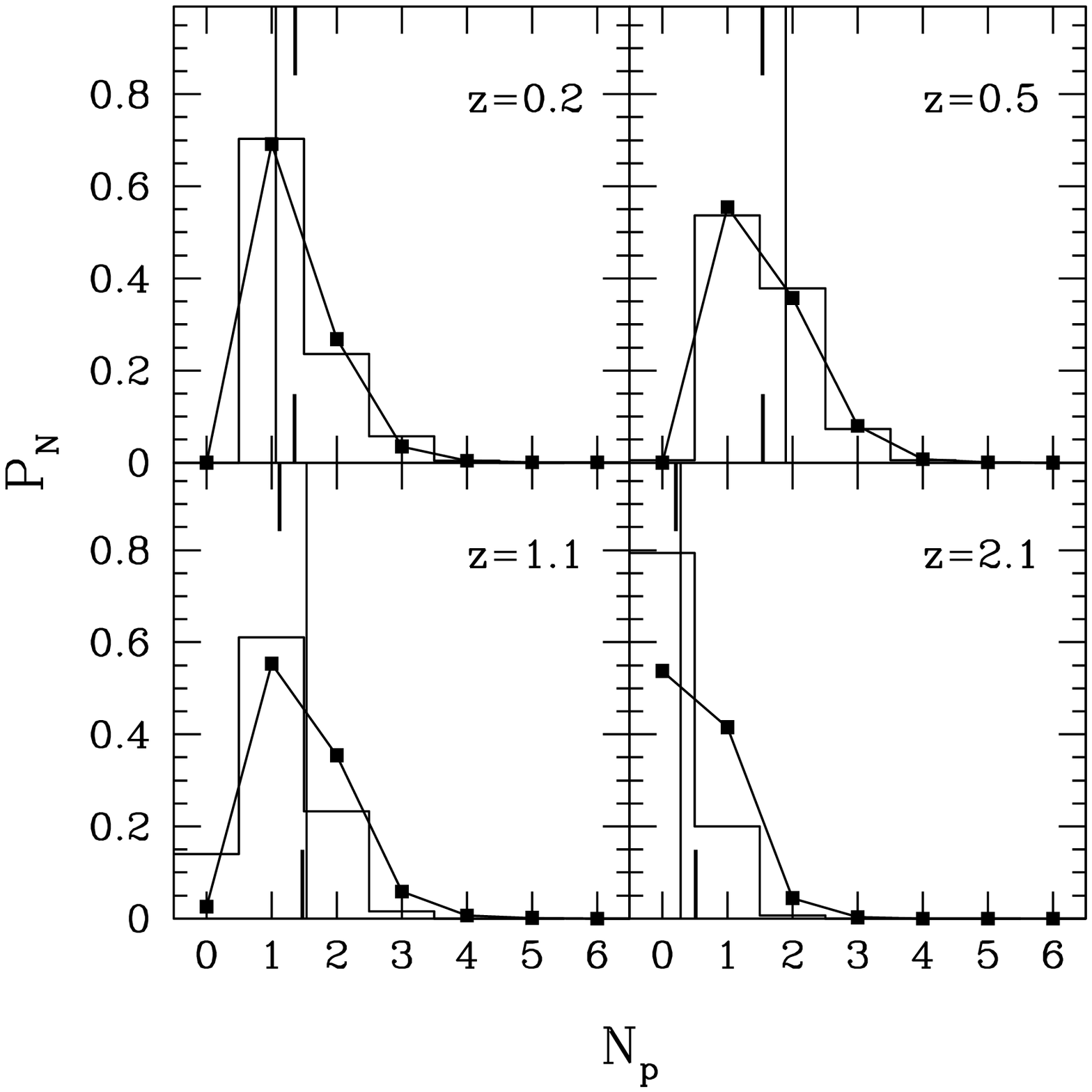,height=8truecm,width=8truecm}}
\centerline{\psfig{file=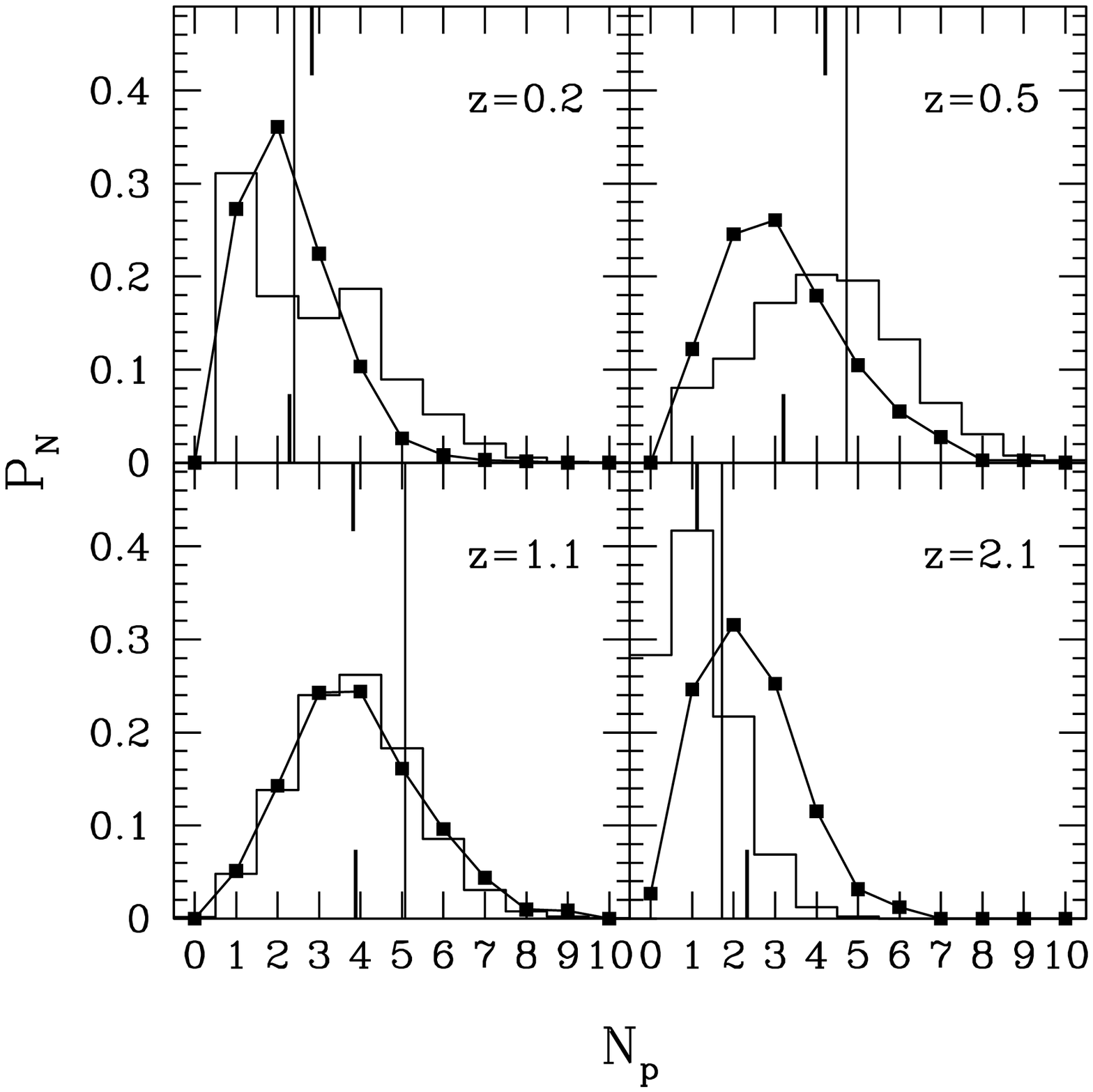,height=8truecm,width=8truecm}}
\caption{The probability that a halo $M_0$ at $z=0$ had $n$ progenitors, at
redshifts 0.2, 0.5, 1.1, and 2.1. The square symbols show the results from the
simulations, and the histogram shows the SAM-trees. The full vertical line
shows the mean predicted by the EPS model. The short lines on the bottom of the
boxes show the means of the simulation distributions, and the short lines on
the top of the boxes show the means for the SAM-trees. top panel: $\bar{M}_0=10
M_l$ bottom panel: $\bar{M}_0=40 M_l$}
\label{fig:pn}
\end{figure}

The ``trunk'' plots discussed in the previous section provide a qualitative
impression of the evolution of the distribution we are interested in, but they
are somewhat difficult to interpret quantitatively. Note that the conditional
mass function shown in \S\ref{sec:comp:means}, $P(M_p)$, is a projection of the
distribution $P(n, M_p)$: $P(M_p) = \sum_{n} P(n, M_p)$. We can also look at
the orthogonal projection, i.e. the probability of having $n$ progenitors,
regardless of their mass, $P(n) = \int_{M_l}^{M_0} P(n, M_p) dM_p$. We show
this quantity for the simulations and the SAM-trees in figure~\ref{fig:pn}. 

The agreement of the distributions in the smaller mass bin, $\bar{M}_0=10\,
M_l$ is quite good. For the larger mass bin, $\bar{M}_0=40\, M_l$, the trees
are skewed towards \emph{larger} numbers of progenitors at $z=0.5$, agree
almost perfectly with the simulations at $z=1$, and are skewed in the opposite
direction (towards smaller number of progenitors) at $z=2$. The sense of these
discrepancies and the reversal of the trend with redshift should be quite
familiar by now and are a direct result of the discrepancies between the EPS
model and the simulations that we have already remarked upon.

\begin{figure}
\centerline{\psfig{file=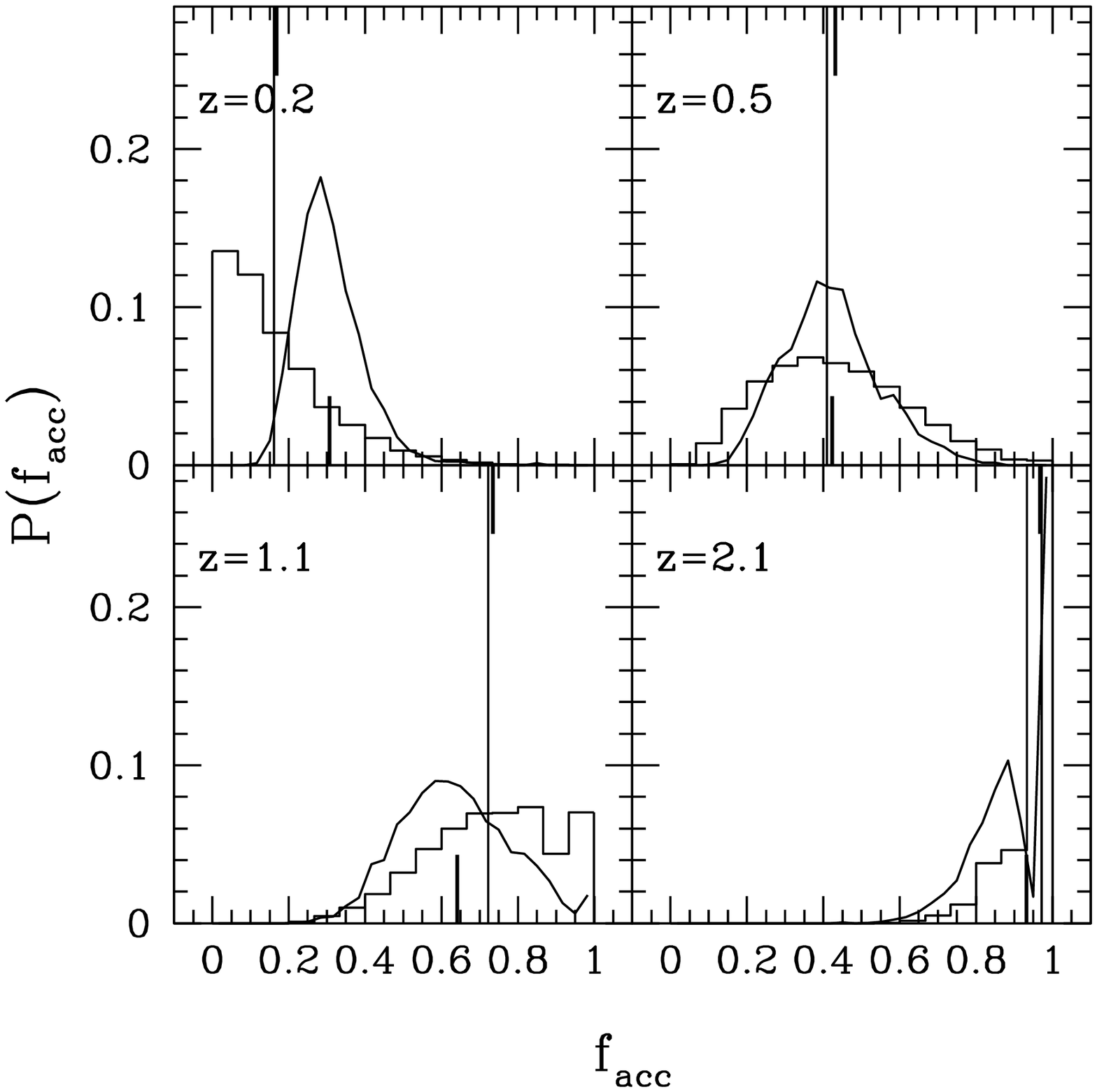,height=8truecm,width=8truecm}}
\centerline{\psfig{file=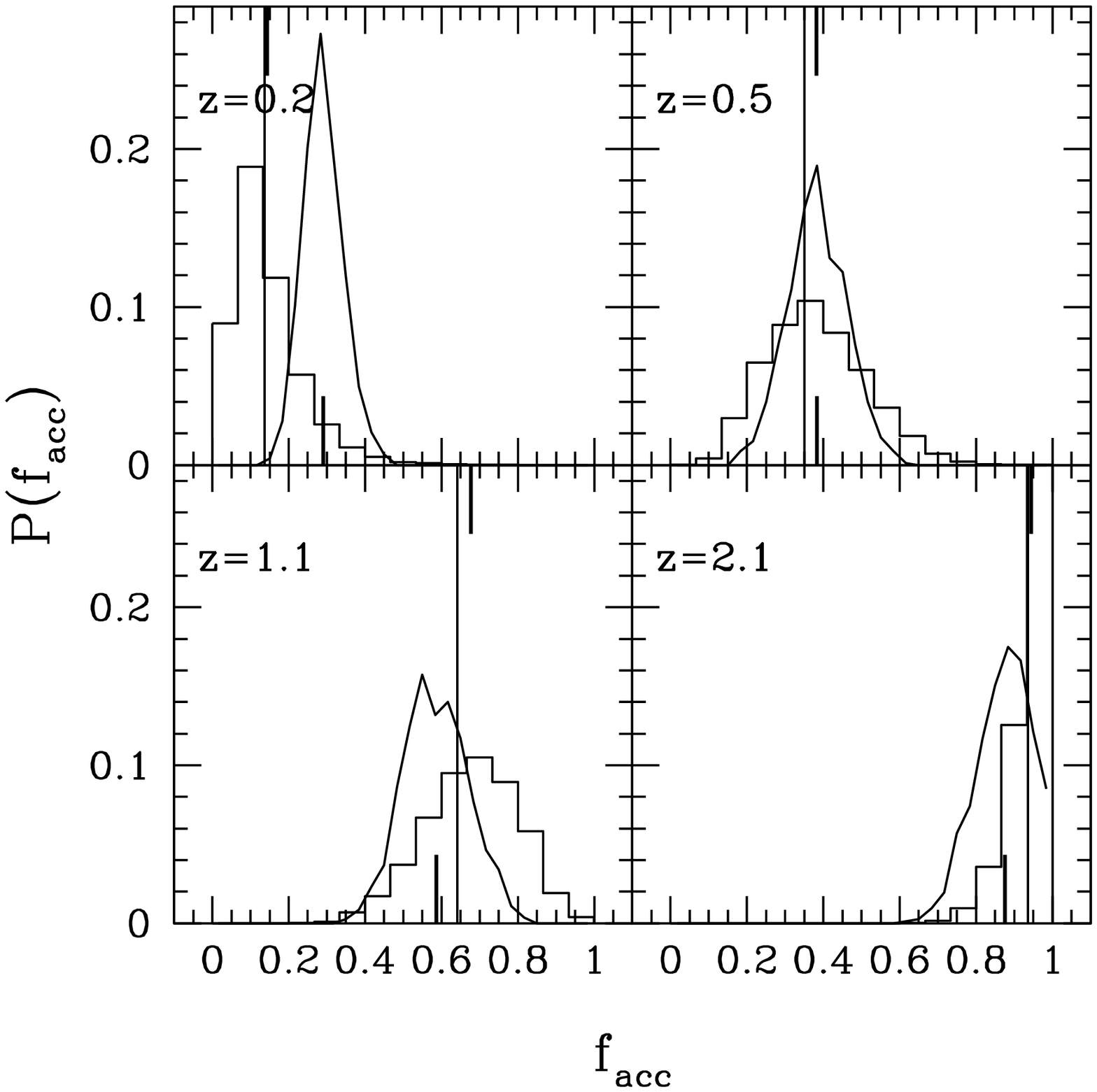,height=8truecm,width=8truecm}}
\caption{The probability that a fraction $f_{\rm acc}$ of the parent mass was
in the form of accreted mass, at redshifts 0.2, 0.5, 1.1, and 2.1. The curves
show the results from the simulations, and the histograms show the
SAM-trees. The full vertical line shows the mean predicted by the EPS
model. The short lines on the bottom of the boxes show the means of the
simulation distributions (inclusive and total), and the short line on the top
of the boxes shows the mean for the SAM-trees. top panel: $\bar{M}_0=10 M_l$;
bottom panel: $\bar{M}_0=40 M_l$}
\label{fig:pfacc}
\end{figure}
Similarly, the projection of the distribution $P(n, f_{\rm acc})$ gives the
probability that a fraction of the parent mass $f_{\rm acc}$ was in the form of
accreted mass, regardless of the number of progenitors. This is shown in
figure~\ref{fig:pfacc}. Once again we can compare the EPS prediction for the
mean value of $f_{\rm acc}$, from equation~\ref{eqn:faccbar}, and shown on the
figures by the full vertical line, with the means of the distributions in the
simulations and the SAM-trees. The mean accreted mass in the simulations is
larger than the EPS prediction at small redshift and smaller at higher
redshift. This is once again a consequence of the original discrepancy
mentioned above.

\subsection{Moments}
\label{sec:comp:moments}
\begin{figure}
\centerline{\psfig{file=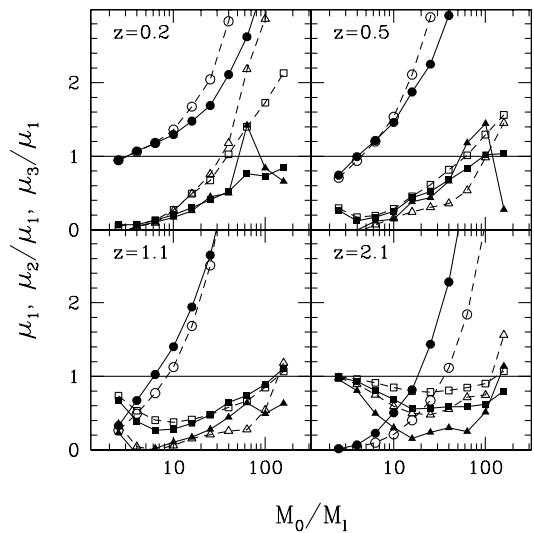,height=8truecm,width=8truecm}}
\caption{Moments of the distribution of the number of progenitors $P_n$, as a
function of parent halo mass $M_0$, for redshifts 0.2, 0.5, 1.1, and
2.1. Dashed lines with open symbols show the results from the SAM-trees, and
solid lines with filled symbols show the results from the simulations. The dots
indicate the first moment (mean), the squares the rescaled second moments (the
usual second moment divided by the first moment), and the triangles the
rescaled third moments (the usual third moment divided by the first
moment). The rescaled moments would be equal to unity for a Poisson
distribution (indicated by the horizontal line). }
\label{fig:pn_moments}
\end{figure}
In figure~\ref{fig:pn_moments}, we show the first moment and rescaled second
and third moments of the distribution of the number of progenitors $P_n$ as a
function of the parent halo mass, $M_0$. The rescaled second and third moments
are the usual second and third moments divided by the first moment, such that
they would be equal to unity for a Poisson distribution. Bold lines are for the
SAM-trees and thin lines are for the simulations. In the first redshift step
($z=0.2$), the moments agree at low masses ($M_0 \approx M_l$), but diverge for
larger $M_0$. The divergence is in the sense that the trees have a larger
first, second, and third moment than the simulations. In the next redshift
step, we can see the hint of the beginning of the reversal of the sense of the
discrepancy, at the low $M_0$ end. The distribution from the SAM-trees is still
somewhat broader (larger second moment) but now is less skewed (smaller third
moment) than the simulations.  The moments seem to cross on all mass scales at
about $z\sim1$ and by $z=2$ the mean number of progenitors is \emph{smaller}
for the SAM-trees on all mass scales. The second and third moments are still
larger in the trees than in the simulations.

\subsection{Largest Progenitors}
\label{sec:comp:lp}
\begin{figure}
\centerline{\psfig{file=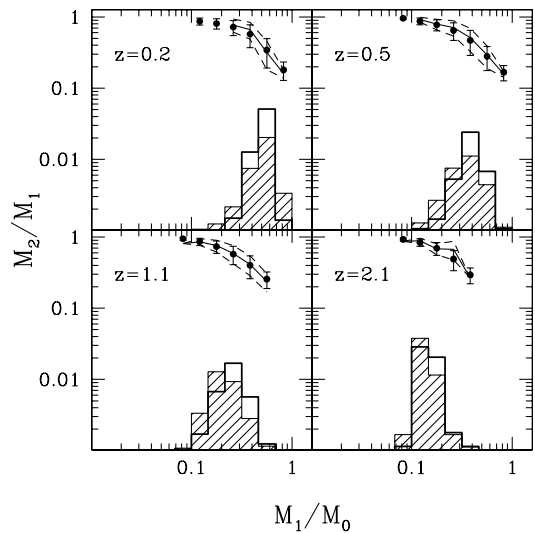,height=8truecm,width=8truecm}}
\centerline{\psfig{file=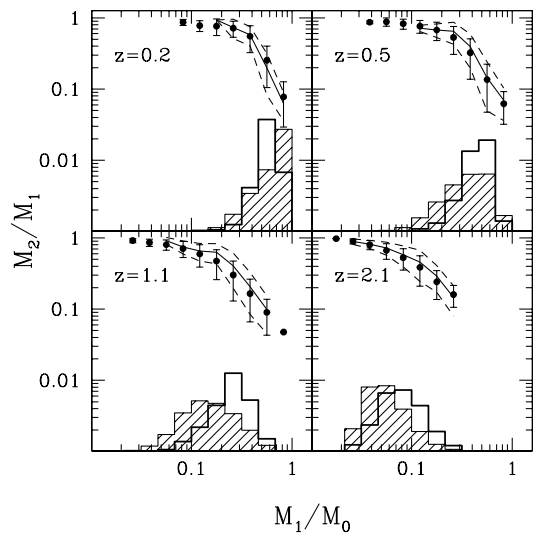,height=8truecm,width=8truecm}}
\caption{Masses of largest progenitors, for redshifts 0.2, 0.5, 1.1, and
2.1. The quantity plotted in the upper part of the figure is the mass ratio of
the second largest to the first largest progenitor ($M_2/M_1$) vs. the ratio of
the largest progenitor to the parent mass ($M_1/M_0$). Solid lines show the
results from the simulations, and filled dots show the results from the
SAM-trees. Dashed lines show the 1-$\sigma$ variance in the simulations, and
the error bars show the 1-$\sigma$ variance in the SAM-trees. The histograms
show the distribution of the mass of the largest progenitor. Shaded histograms
are for the SAM-trees, and unshaded histograms are for the simulations.  }
\label{fig:lp}
\end{figure}
A statistic which should be important in determining the properties of the
galaxies that form within dark matter halos is the mass ratio of the largest
and second largest progenitors. Figure~\ref{fig:lp} shows the ratio of the
second largest progenitor to the largest progenitor $M_2/M_1$, as a function of
the ratio of the mass of the largest progenitor to the parent, $M_1/M_0$. We
find very good agreement between the SAM-trees and the simulations in both the
mean and the variance of this quantity. Apparently, the usual discrepancy in
the progenitor mass distributions cancels out. As we see from the histograms at
the bottom of the figure, the mass of the largest progenitor is \emph{larger}
in the SAM-trees than in the simulations at low redshift, and \emph{smaller} at
higher redshift. This clearly follows from the original effect shown in
figure~\ref{fig:cmf}.

\section{Discussion and Conclusions}
\label{sec:conclusions}
Many of the previous comparisons of the Press-Schechter model with simulations
have emphasized the agreement between the two, which is surprisingly good in
view of the crude assumptions underlying the model. Certainly this model has
served as a useful tool in the study of structure formation. However, as
comparisons of theory and observations become more refined, it is important to
be aware of the limitations of the model, and to quantify its inaccuracies.

There are several implications of the errors associated with the PS and EPS
model, corresponding to various implementations of the formalism. First, the
overall abundance of halos of a given mass may be over- or underestimated by
the PS model compared to $N$-body simulations, depending on the mass scale and
the redshift. The general sense of this discrepancy is that the PS model
\emph{overestimates} the halo abundance on mass scales less than $\sim 10^{14}
\msun$ by about a factor of 1.5 to 2.0 at $z=0$. The model and simulation
results agree well on all mass scales at an intermediate redshift $z\sim1$, but
this agreement is momentary and fortuitous: it appears to be simply the point
where the two functions happen to cross as they evolve in redshift. At higher
redshifts, the abundance of large mass halos $M\ga10^{12}$ is
\emph{underestimated} by the PS model. Because this is the exponentially
decreasing part of the mass function, this error can be quite serious. This
behaviour seems to be qualitatively similar regardless of the cosmology or power
spectrum, although the evolution with redshift (e.g. the redshift of fortuitous
agreement) will be somewhat different. The precise magnitude of the discrepancy
between PS and simulations depends on the cosmology, power spectrum, and the
implementation of the Press-Schechter model (i.e. top-hat vs. Gaussian
smoothing, value of $\delta_{c,0}$, etc.), but the dependence on mass-scale and
redshift implies that the discrepancy cannot be removed by simply adjusting the
parameter $\delta_{c,0}$ (see also \citeNP{gsphk:98}).

The second point is that this discrepancy is echoed qualitatively in the
comparison of the conditional mass function predicted by the EPS model with the
simulations. This in turn affects the predictions of merger rates, accretion
rates, formation times, and the distribution of progenitor masses in merger
trees based on the EPS formalism. We have shown that the distribution of the
number of progenitors found in the merger trees is in fairly good agreement
with the results of the simulations. On galaxy scales ($M\sim10^{12}\msun$)
these quantities are in excellent agreement. On larger mass scales
($M\sim10^{13}\msun$, about the size of a group), the number of progenitors is
over-estimated by the SAM-trees by at most a factor of two. The discrepancy
increases further as the mass of the parent halo increases. This effect
artificially steepens the faint-end slope of the luminosity function in
EPS-based semi-analytic models of galaxy formation (e.g. \citeN{kwg},
\citeN{cafnz}, \citeN{sp:98}). This contributes to an apparent discrepancy
with the observed slope, which seems to require some mechanism that strongly
suppresses star formation in small-mass objects, such as strong supernovae
feedback. The error resulting from the Press-Schechter model does not eliminate
the need for such a mechanism but it may mean that milder, more realistic
feedback will prove to be sufficient.

The two-dimensional distribution of progenitor mass and number and its
evolution with redshift show the same qualitative behaviour in the SAM-trees
and the simulations, although there are differences resulting directly from the
discrepancies mentioned above. The second and third moments of the distribution
of the number of progenitors agree fairly well, to better than a factor of two
in the worst case examined here. The agreement between the first, second, and
third moments of the SAM-tree and simulation distributions diverges
systematically as the parent halo mass increases. The variance in the SAM-tree
distributions is always larger than in the simulations. The mass ratio of the
largest two progenitors, and the variance in this quantity, are in good
agreement in the SAM-trees and simulations.

We have focussed on only one implementation scheme for building merger trees
from the EPS theory (SK). Different schemes may yield different distribution
functions for the various quantities we have presented here; however, we have
argued that most of the important effects that we have discussed can be traced
directly to the original EPS model and are not specific features of the merging
tree scheme. We note that our conclusions are very consistent with the findings
of
\citeN{kauffmann:98}, who compared galaxy formation models based on a different
merger-tree method \cite{kw} with models implemented directly within N-body
simulations (the same simulations analyzed here). They note the same overall
factor of two discrepancy in the halo mass function, but find that the
luminosity function for a halo of a given mass is very similar in the two
cases. This is presumably because, as we have noted, the relative properties of
the progenitors for a halo of a given mass are in good agreement in the
SAM-trees and the simulations.

Our main conclusions from this study may be summarized as follows: 
\begin{itemize}
\item The Press-Schechter and Extended Press-Schechter models exhibit
discrepancies with N-body simulations at the level of up to 50 percent. This can
lead to errors in the mean numbers and masses of progenitors and the higher
order moments of the progenitor number-mass distributions as well as the overall
and conditional halo mass functions. More serious errors may arise at high
redshifts for the rare objects at the exponentially declining high-mass end of
the mass distribution (clusters). These effects must be taken into account in
any detailed comparison of PS or EPS based models with observations.

\item Despite these problems, {\it relative} properties of the progenitor distribution
(such as the progenitor mass ratios) for a halo of a given mass are quite well
reproduced in the merger trees, especially on galaxy mass scales. Halo merging
histories constructed using the EPS formalism should thus provide a reasonably
reliable framework for semi-analytic modelling of galaxy formation.
\end{itemize}

\section*{Acknowledgments}
The $N$-body simulations analysed in this paper were carried out by the Virgo
Supercomputing Consortium (http://star-www.dur.ac.uk/~frazerp/virgo/virgo.html)
using computers based at the Max--Planck--Institut f\"ur Astrophysik, Garching
and the Edinburgh Parallel Computing Centre. We thank J\"org Colberg for a
thorough reading of the text and helpful comments. RSS acknowledges support from
an NSF GAANN fellowship. This work was supported in part by NASA ATP grant
NAG5-3061 at the University of California, Santa Cruz, and by the US-Israel
Binational Science Foundation grants 95-00330 and the Israel Science Foundation
grant 950/95.

\bibliographystyle{mnras}
\bibliography{mnrasmnemonic,trees}

\end{document}